\documentclass{iaus}
\usepackage{graphicx}

\title[PP 86.~~Exoplanet Mapping Revealed] 
{Inverting Phase Curves to Map Exoplanets}

\author[Cowan \& Agol]   
{Nicolas B. Cowan$^1$
 \and Eric Agol$^1$}

\affiliation{$^1$Astronomy Department, University of Washington,
   Box 351580, Seattle, WA  98195 \\email: {\tt cowan@astro.washington.edu, agol@astro.washington.edu}}

\pubyear{2008}
\volume{253}  
\pagerange{??}
\jname{Transiting Planets}
\editors{A.C. Editor, B.D. Editor \& C.E. Editor, eds.}
\begin{document}

\maketitle

\begin{abstract}
One of the most exciting results of the Spitzer era has been the ability to construct longitudinal brightness maps from the infrared phase variations of hot Jupiters. We presented the first such map in \cite[Knutson et al. (2007)]{Knutson_2007}, described the mapping theory and some important consequences in \cite[Cowan \& Agol (2008)]{Cowan_2007} and presented the first multi waveband map in \cite[Knutson et al. (2008)]{Knutson_2008}. In these proceedings, we begin by putting these maps in historical context, then briefly describe the mapping formalism. We then summarize the differences between the complementary N-Slice and Sinusoidal models and end with some of the more important and surprising lessons to be learned from a careful analytic study of the mapping problem.
\end{abstract}

\firstsection 
\section{Introduction}
Observations of secondary eclipses in exoplanetary systems, starting with HD~209458b \cite[(Deming et al. 2005)]{Deming_2005} and TrES-1b \cite[(Charbonneau et al. 2005)]{Charbonneau_2005}, made it possible to estimate the integrated day-side brightness of transiting exoplanets. Constraining the \emph{global} brightness map of exoplanets, on the other hand, requires observations at various orbital phases, involving more sophisticated calibration of observations, much longer observing campaigns, or both. The first measurements of thermal phase curves for exoplanet systems were reported by \cite[Harrington et al. (2006)]{Harrington_2006}, which reported a large phase function for $\upsilon$ Andromeda~b, and \cite[Cowan, Agol \& Charbonneau (2007)]{Cowan_2007}, which detected a phase function for HD~179949b, and obtained useful upper limits for HD~209458b and 51~Peg~b. These results proved valuable in constraining the day-night brightness contrast ---and hence the energy recirculation efficiency--- of those planets and indicated that hot Jupiters represent a heterogeneous group.
Those first two studies, however, had very incomplete phase coverage (5 epochs for the  \cite[Harrington et al. 2006]{Harrington_2006} campaign, and 8 epochs for each of the \cite[Cowan et al. 2007]{Cowan_2007} campaigns). Furthermore, three of the four observed planets were not in transiting systems, and the one transiting system (HD~209458) was deliberately observed outside of transit or secondary eclipse.

The $33$~hours of continuous monitoring of HD~189733b presented in \cite[Knutson et al. (2007)]{Knutson_2007} differs in three important ways from those first detections of phase variations: 1) The observed system exhibits transits, so the planet's orbital inclination with respect to the celestial plane is known. 2) A secondary eclipse of the planet was observed during the course of the observations, making it possible to quantify not just the relative but the \emph{absolute} flux of the planet as a function of orbital phase. 3) The continuous observing campaign, the system's relative proximity to the Earth, its favorable contrast ratio, and ingenious corrections for detector systematics conspired to produce the highest S/N light curve of its kind ever measured. Although the observations spanned little more than half an orbit of HD~189733b, the unprecedented quality of the light curve enabled us not only to measure the planet's day/night contrast, but also to generate the first ever brightness map of an extrasolar planet.

\begin{figure}[htb]
\includegraphics[width=0.5 \textwidth]{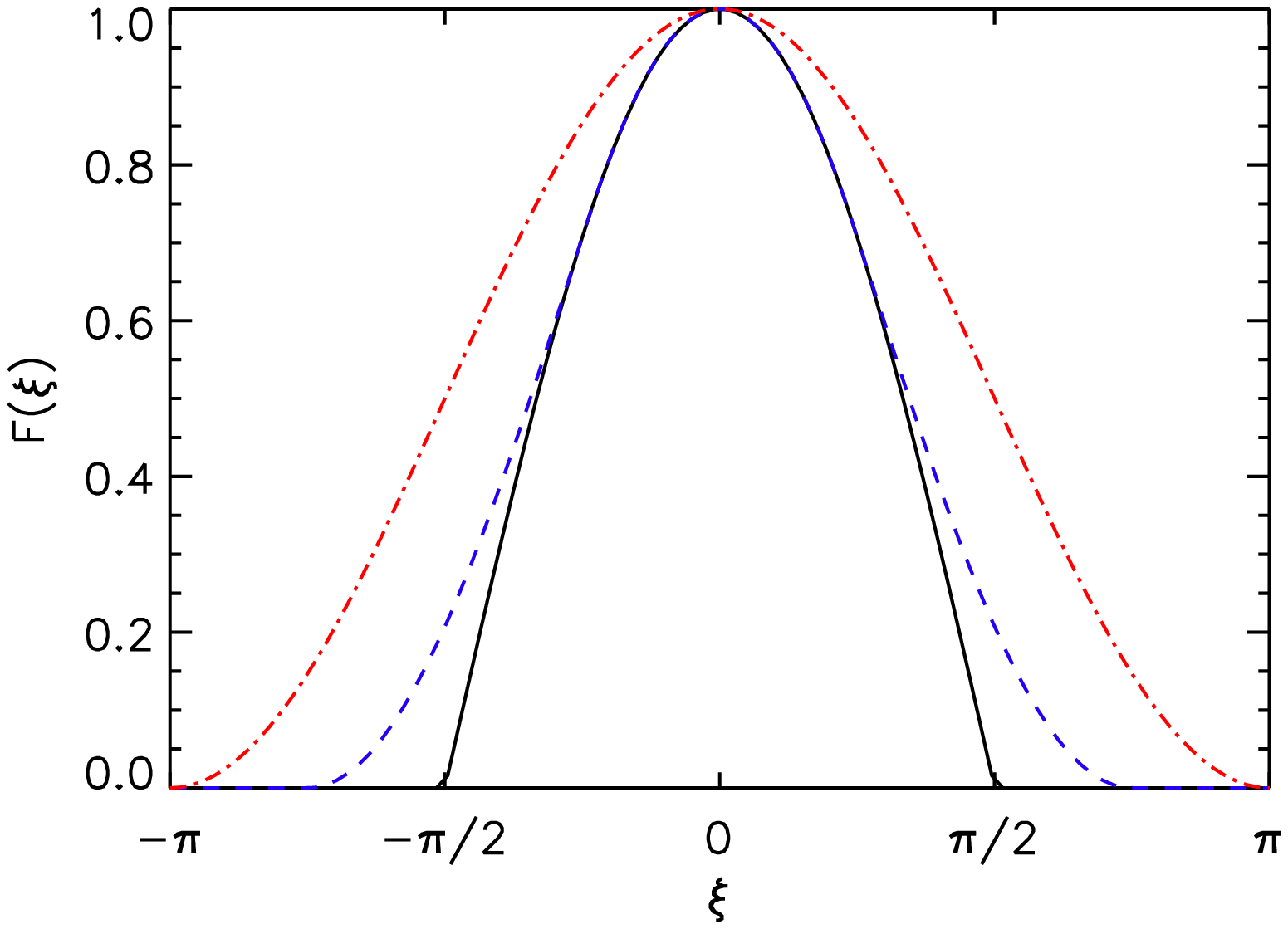} \includegraphics[width=0.5 \textwidth]{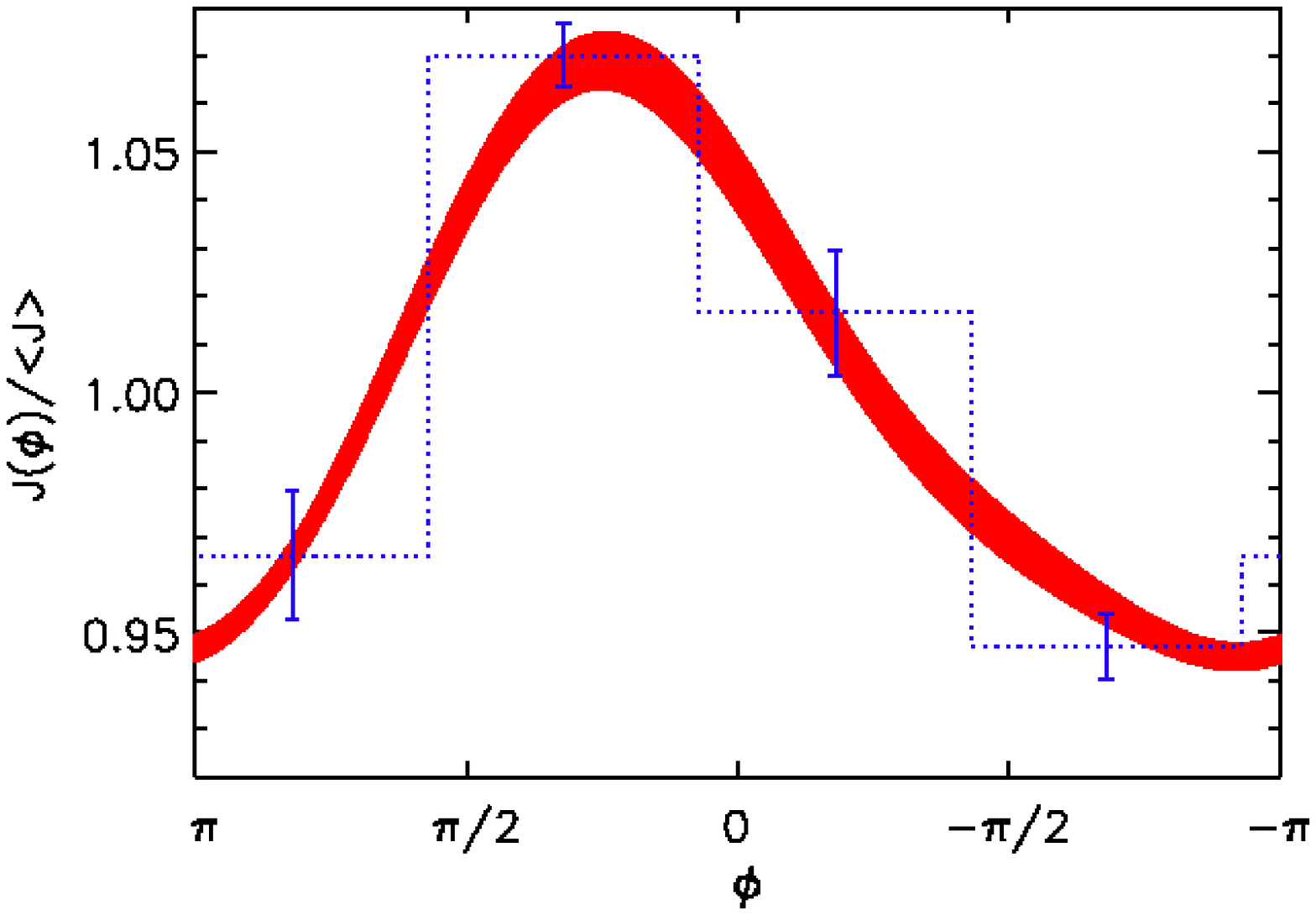}
\caption{The solid line in the left panel shows the phase function response to a $\delta$-function in brightness (AKA ``the kernel''). The dashed and and dot-dashed lines represent the contributions from a single slice in 4-slice and 2-slice models, respectively. In the right panel, the solid band shows the 1-$\sigma$ range for a sample sinusoidal map, while the dotted line shows the equivalent N-slice map and associated error bars. Both maps have 5 parameters and produce indistinguishable light curves.}
\label{slice_kernels}
\end{figure}

\section{Necessary and Sufficient Assumptions for Mapping}
There are three necessary and sufficient conditions for phase function mapping to be feasible (see also the orginal formalism of \cite[Russell, 1906]{Russell_1906}):
\begin{enumerate}
\item One must be able to remove from the observed light curve any stellar variability (eg: star spots rotating into and out of view) as well as detector systematics (detector ramps, intra-pixel sensitivity, etc.). \cite[Knutson et al. (2008)]{Knutson_2008} presents the most sophisticated treatment of these effects to date.
\item One must neglect limb darkening in the planet's atmosphere.  This is a reasonable approximation at mid-IR wavelengths, leading to errors of less than 1\% \cite[(Cowan \& Agol 2008)]{Cowan_2008}.
\item One must assume that the large-scale weather of the planet is in a steady-state.  This means that the global hot spots, cold spots and jet streams do not vary in brightness or shift with respect to the substellar point over a single planetary orbit. This assumption appears to hold at the 5--10\% level for hot Jupiters on circular orbits (Agol et al. in prep).  
\end{enumerate}

\section{Planet Mapping Formalism}
For an edge-on system we define the phase angle $\xi$, which corresponds to the observer--planet--star angle ($\xi = 0$ at secondary eclipse; $\xi=\pi$ at transit), as well as the longitude, $\phi$, and latitude, $\theta$, in a rotating frame, such that $\phi=0$ at the sub-stellar point, $\theta=0$ at the planet's north pole, and $\phi$ increases in the same sense as $\xi$. The condition of a steady-state weather pattern can be expressed as requiring that the specific intensity, $I(\phi,\theta)$, is unchanging with time. 

There are no current observations which can constrain the $\theta$-dependence of $I$, but for edge-on orbits the latitudinal dependence of the intensity is unimportant since one can define $J(\phi) = \int_{0}^{\pi} I(\phi,\theta) \sin^{2}\theta d\theta$, which represents the flux contribution from an infinitesimal slice of the planet when viewed face-on. The flux, $F$, we observe from the planet at a given orbital phase can then be written as a convolution, $F(\xi) = \int_{0}^{2\pi} J(\phi) K(\phi,\xi) d\phi$, with the piece-wise defined kernel, $K(\phi, \xi) = \max\left(\cos(\phi + \xi), 0 \right)$. The kernel represents the response of the phase function to a delta function in $J(\phi)$, and it is very broad, with a full width at half-maximum of $2\pi/3$, as shown by the solid line in the left panel of Figure~\ref{slice_kernels}.    

\section{Model Longitudinal Maps}
The convolution described in the previous section transforms a given longitudinal map into an observed light curve.  The more challenging problem if how to reliably de-convolve an observed light curve to obtain the longitudinal map of a planet. In \cite[Cowan \& Agol (2008)]{Cowan_2008} we presented two complementary models (examples of which are shown in the right panel of Figure~\ref{slice_kernels}), described below.

{\bf N-Slice Model:} This model consists of equal-size longitudinal slices of uniform brightness (think beach ball). Such maps simplify the convolution, enabling the use of brute force numerical techniques (least-squares, MCMC, etc.) to determine the best-fit longitudinal map given an observed light curve. This approach is versatile, easily adapted to non-transiting planets or planets with incomplete light curves. Although an N-slice longitudinal map is neither differentiable nor realistic, smoothing the map does not significantly change the resulting light curve.  On the other hand, the brightness of the different slices do not depend on the light curve in a linearly independent fashion, so using too many slices to model a light curve with poor S/N makes the uncertainty in \emph{all} of the slices blow up. 

{\bf Sinusoidal Model:} Sinusoids are orthogonal eigenfunctions of the convolution described in \S~3. An observed light curve can therefore be decomposed via a Fourier expansion, then trivially transformed into a sinusoidal map using Equation~7 of \cite[Cowan \& Agol (2008)]{Cowan_2008}. Sinusoidal longitudinal maps have the advantage of being imminently believable, but for incomplete phase curves the uncertainty in the map does not have have the properties one would like. For example, if a phase function is only obtained for half of a planetary orbit, the uncertainty in the map is no greater for the hemisphere which was not well observed. Fortunately, Warm Spitzer's propensity for longer observing campaigns will be perfectly suited for obtaining full phase curves \cite[(Deming et al. 2007)]{Deming_2007}.   

\begin{figure}[htb]
\includegraphics[width=0.5\textwidth]{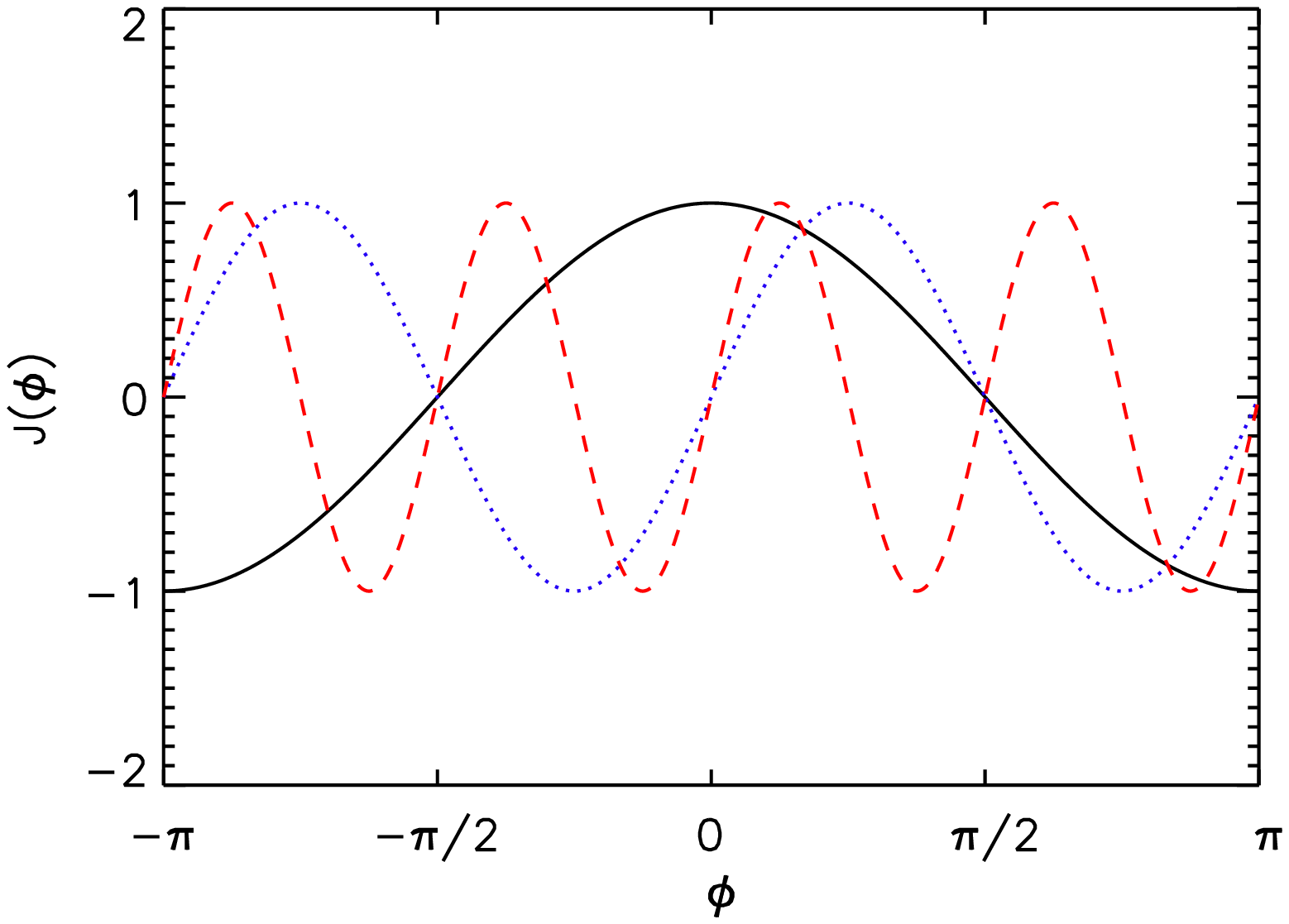} \includegraphics[width=0.5 \textwidth]{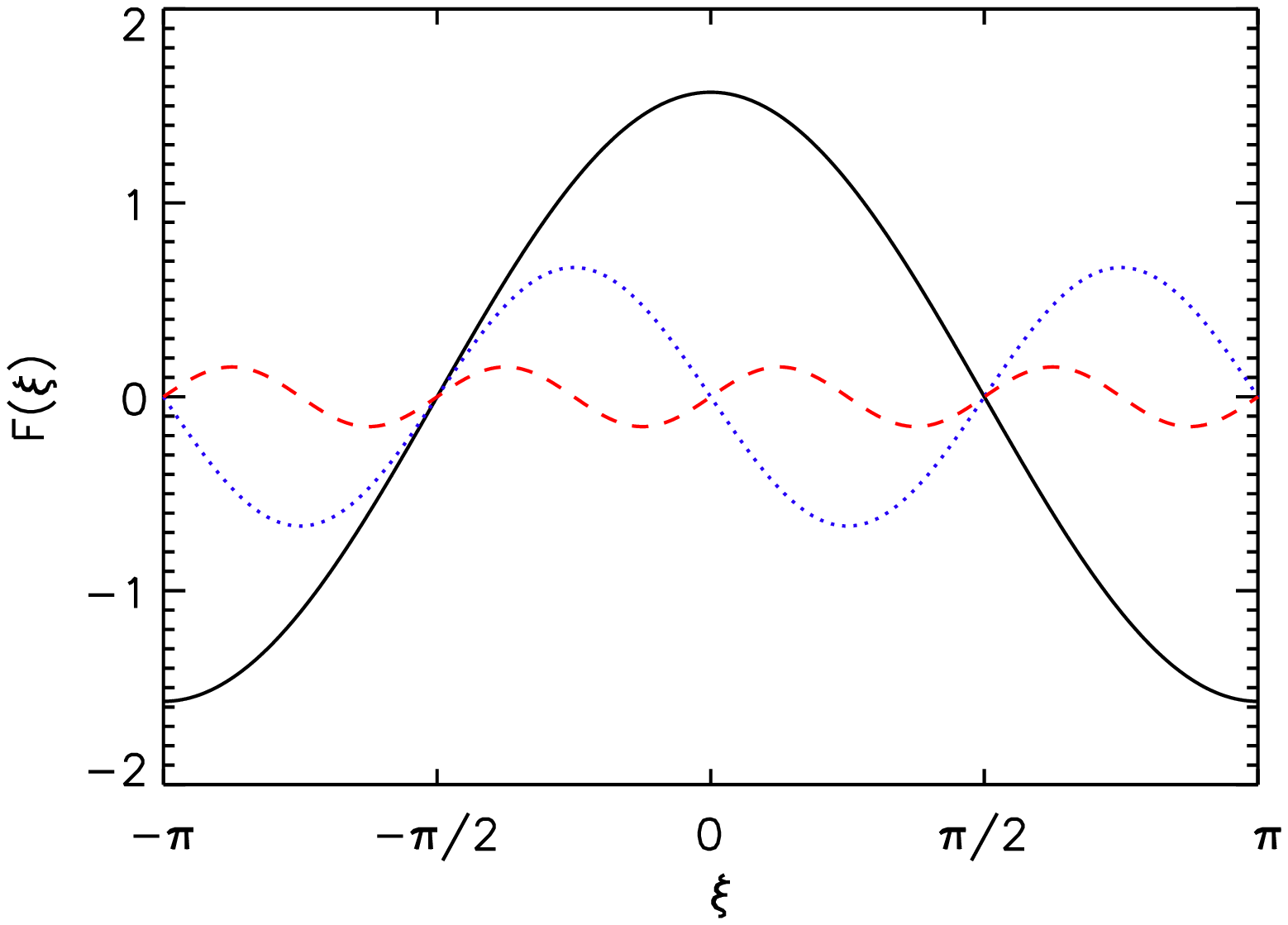}
\caption{The left panel shows the $j=1$, $j=2$ and $j=4$ sinusoidal maps, while the right panel shows the resulting phase variations. The higher-frequency modes are damped out because a full hemisphere is visible at any point in time.}
\label{sinusoidal_maps}
\end{figure}

\section{General Mapping Considerations}
\begin{figure}[htb]
\includegraphics[width=0.5 \textwidth]{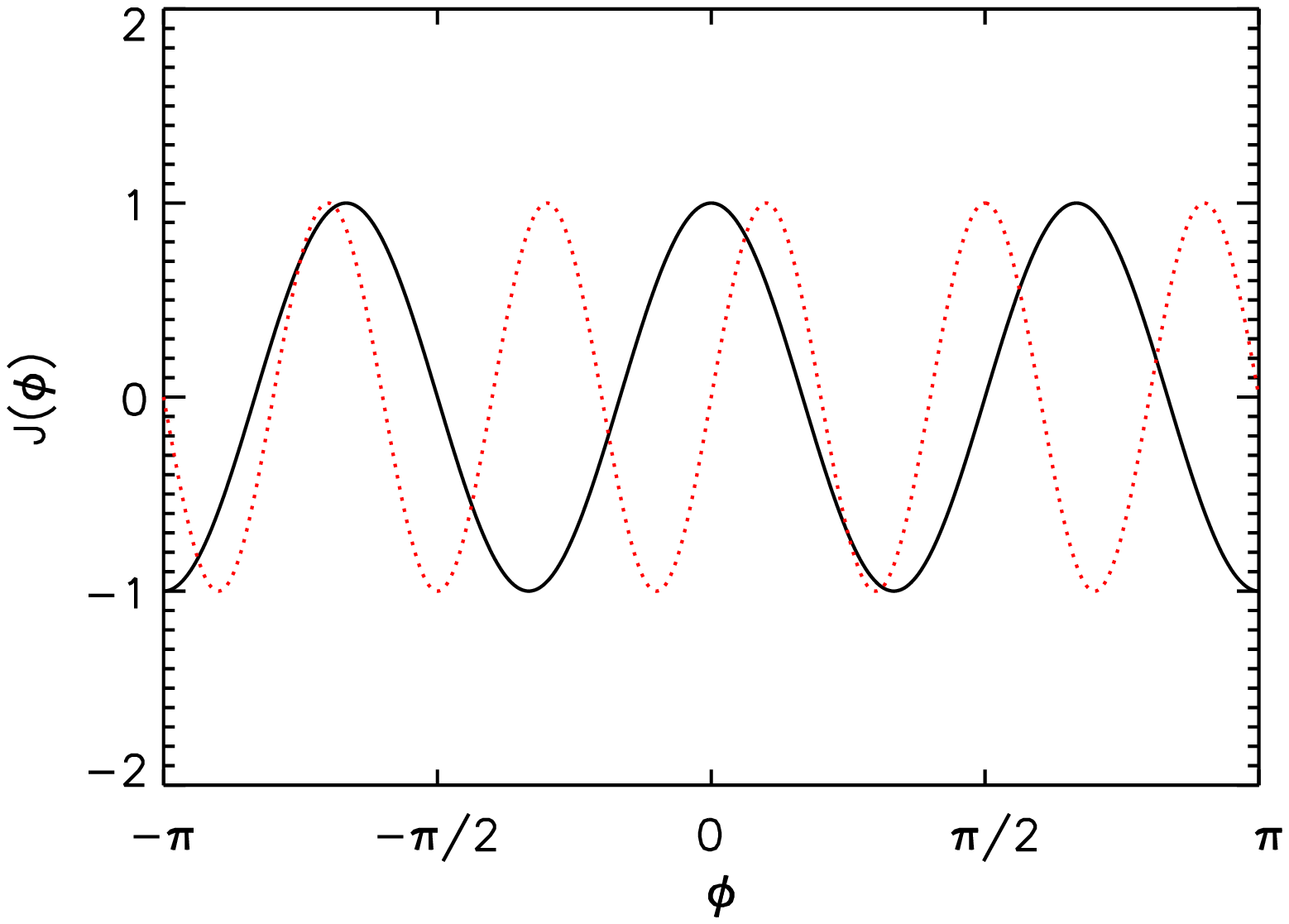} \includegraphics[width=0.5 \textwidth]{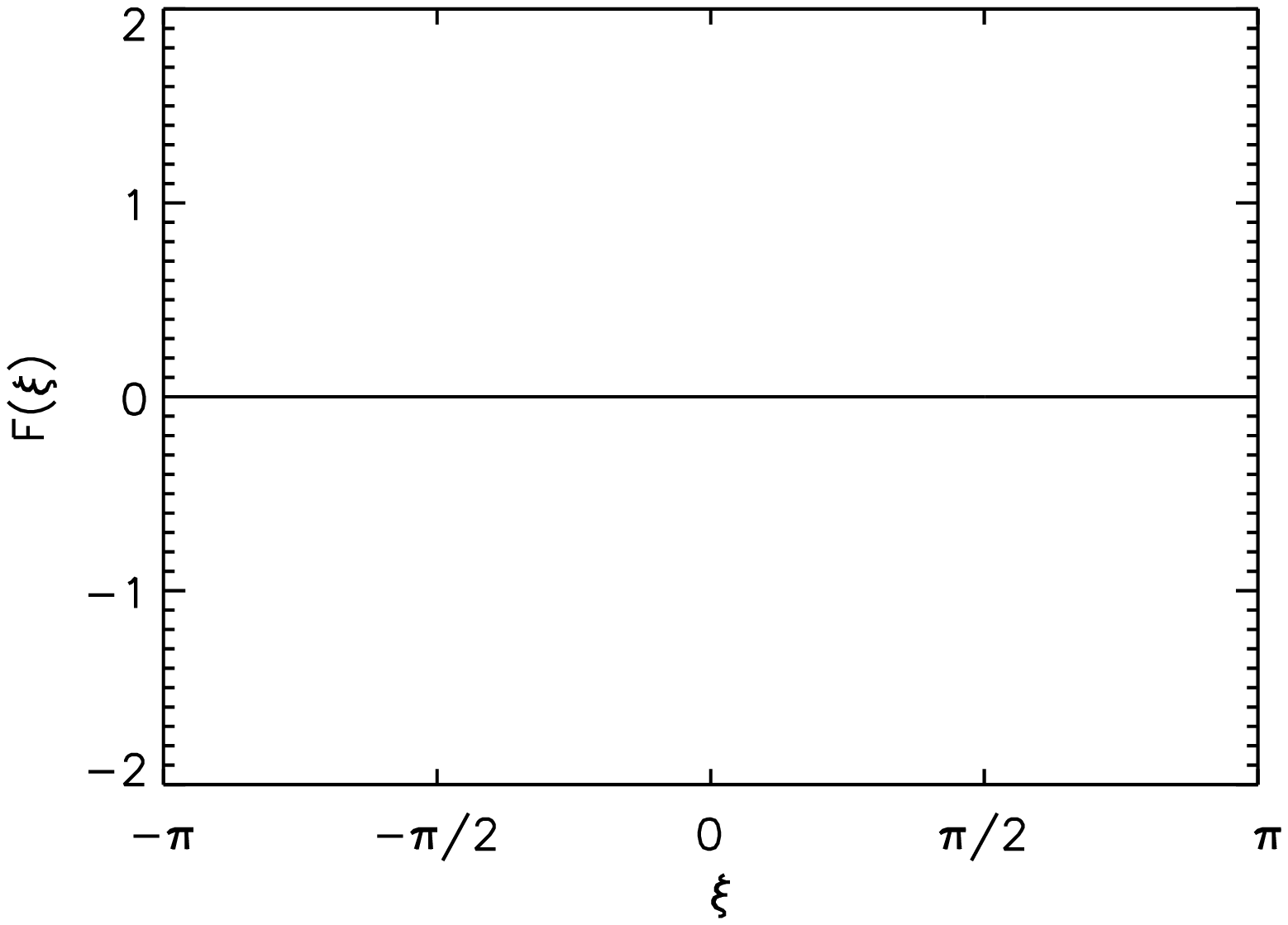}
\caption{The left panel shows $j=3$ and $j=5$ sinusoidal maps, while the right panel shows the resulting phase variations. All odd sinusoidal maps above $j=1$ cancel out by symmetry and are therefore invisible to observers.}
\label{odd_sinusoidal_maps}
\end{figure}

The Sinusoidal Model provides an instructive tool for studying the mapping problem, since the maps and associated light curves are simple analytic functions. Figure~\ref{sinusoidal_maps} shows how the smoothing kernel suppresses high-frequency spatial brightness variations.  This is a direct consequence of seeing half of the planet at a time.  Technically, one only sees $1/3$ of the planet particularly well at any point in time (recall the $2\pi/3$ FWHM). This leads to the pernicious problem shown in figure Figure~\ref{odd_sinusoidal_maps}: the kernel entirely wipes out odd sinusoidal modes (except for $j=1$). In other words, if a planet's dominant weather consisted of three equally spaced hot spots near its equator, it would exhibit \emph{no} phase variations!

The invisibility of odd modes is not merely an intellectual curiosity: it sets a hard limit on the accuracy of longitudinal maps.  If the $j=3$ modes in the planet's longitudinal brightness profile are not visible, there is not much to be gained by extending the Fourier expansion to $j=4, 6$, etc.  Those modes may well be precisely measured, but this will do nothing to increase the \emph{accuracy} of the resulting planet map. To flip this problem on its head, a simple way to test the assumptions of \S~2 is to look for $j=3$ modes in the observed light curve.  The bottom line is that one can do no better than a second-order Fourier expansion of an observed light curve: $F(\xi) = F_{0}+ F_{1}\cos\left(\xi-\xi_{1}\right) + F_{2}\cos\left(2(\xi-\xi_{2})\right)$.  By the same token, a limit of 5 free parameters (4 slices $+$ a phase offset, or just 5 slices) applies to the N-Slice maps.  Maps with many more parameters than this can be made, but should be treated with skepticism.

\end{document}